%Paper: hep-ph/9302314
%From: Jutta Kunz <kunz@fys.ruu.nl>
%Date: Fri, 26 Feb 1993 16:27:33 +0100 (MET)

\magnification 1200
\baselineskip=16pt
\parskip 10pt plus 1pt minus 1pt
\hsize 15 true cm \vsize 23 true cm
\overfullrule 0pt
\ \
\vskip 1 cm
\centerline {\bf{SPHALERONS AT FINITE TEMPERATURE}}
\vskip 2.0  true cm
\centerline {\bf {Sylvie Braibant, Yves Brihaye}}
\centerline {Facult\'e des Sciences, Universit\'e de Mons-Hainaut}
\centerline {B-7000 Mons, Belgium}

\centerline {\bf {Jutta Kunz}}
\centerline {Instituut voor Theoretische Fysica,
             Rijksuniversiteit te Utrecht}
\centerline {NL-3508 TA Utrecht, The Netherlands}
\centerline {and}
\centerline {FB Physik, Universit\"at Oldenburg, Postfach 2503}

\centerline {D-2900 Oldenburg, Germany}

\vskip 3 true cm
\centerline {\bf Abstract}
\noindent

We construct the sphaleron for several
temperature dependent effective potentials.
We determine the sphaleron energy as a function of temperature
and demonstrate that
the sphaleron energy at a given temperature $T$ is well
approximated by the sphaleron energy at temperature zero
scaled by the ratio of the vacuum expectation values
of the Higgs field at temperatures $T$ and zero.
We address the cosmological upper bound on the Higgs mass.

\vfill   \noindent Univ. Mons-Hainaut-Preprint \hfill \break
         \noindent Univ. Utrecht Preprint THU-93/01\hfill \break
         \noindent January  1993
\vfill\eject

{\bf Introduction}

\par The observation [1]
that the baryon asymmetry of the universe (BAU) might possibly
be explained in the framework of the standard model
attracted a lot of attention [2-7].
While in previous scenarios the BAU was produced at high temperature,
e.~g.~in GUTs, Kuzmin et al.~[1] realized, that the BAU might be
produced during the electroweak phase transition.

When, on the one hand,
one considers a scenario for baryogenesis,
where the baryon asymmetry is produced
by some mechanism at high temperature
one has to require that the presently observed BAU
survives the electroweak phase transition.
When, on the other hand,
one considers a scenario for baryogenesis,
where the BAU is generated during the electroweak
phase transition,
one must require that the baryon number violating transitions
have a large enough rate.
In both cases one needs to evaluate the rate of baryon number
violation in the broken phase of the electroweak model.
(See [6,7] for reviews.)

When the temperature cools down, the universe goes through
the electroweak phase transition,
where the symmetry is broken by the Higgs potential.
In the broken phase, the structure of the vacuum
becomes non-trivial. Topologically distinct
vacua are separated by finite energy barriers,
whose height is determined by the sphaleron energy, $E_{\rm sp}$.
The sphaleron [8], an unstable solution of the electroweak model,
plays a central role in the generation of the baryon asymmetry,
since the rate of baryon number violating transitions is largely
determined by the Boltzmann factor
$$  \Gamma \sim  \exp{\Bigl(-{{E_{\rm sp}(T)\over T}}\Bigr)}
\ . \eqno(1)$$

The existence of the BAU might even yield cosmological constraints
for the parameters of the standard model.
Requiring for instance that the BAU, generated
during a first order electroweak phase transition,
does survive till the present time,
means that the baryon number violating transitions
have to be out of thermal equilibrium after
the phase transition.
This constraints the value of the energy barrier,
i.~e.~the sphaleron energy $E_{\rm sp}(T)$,
in the region of temperature where the phase transition occurs.
Shaposhnikov [9] has derived the model-independent relation
$$    { E_{\rm sp}(T_t) \over T_t}  > 45 \ ,   \eqno(2)$$
where $T_t$ denotes the transition temperature.
Relation (2) can be used to obtain
an upper bound on the Higgs mass [2,7,9].

At the moment a major source of uncertainty
in computing the rate of baryon number violating transitions
and in extracting the cosmological upper bound for the Higgs mass
lies in the inclusion of
finite temperature effects in the electroweak model.
Lacking a satisfactory alternative method,
the technique of effective potentials, computed perturbatively
by resumming the dominant Feynman diagrams, is used
to describe the interactions of the standard model in the
neighbourhood of the critical temperature [10].
The simplest temperature dependent effective potential
yields a second order phase transition [6].
Recently several ``improved'' effective potentials
have been proposed [9,11-15], which contain cubic
terms in the Higgs field, providing a first order
electroweak phase transition.
For these temperature dependent effective potentials
one must then reevaluate the sphaleron energy $E_{\rm sp}(T)$
and the transition rate (1) [16].

The purpose of this note is to determine the
sphaleron energy in the region of the phase transition
for three increasingly sophisticated
temperature dependent effective potentials.
Besides computing the sphaleron energy numerically
we employ for the sphaleron energy the simple formula
$$E(\lambda, T) = E(\lambda, T=0)
{\langle \phi (T) \rangle \over {\langle \phi (0) \rangle }}
\ , \eqno(3)$$
where $\langle \phi (T) \rangle$ is the vacuum expectation
value of the Higgs field at temperature $T$.
Comparison of both energies then yields the quality of
the approximation (3) for the more sophisticated
temperature dependent effective potentials.
We also address the model-independent bound (2)
for the sphaleron energy and the implications
for the cosmological upper bound on the Higgs mass
for the potentials considered.

{\bf Sphalerons at zero temperature}

Let us first consider the
classical lagrangian for the electroweak interactions.
It is sufficient
to treat the mixing angle $\theta_w$
perturbatively, as demonstrated recently [17].
In leading order, we therefore
consistently set the U(1) gauge field equal to zero.
The bosonic sector of the electroweak model then reduces to
$$ {\cal L} =  -{1\over 4} F_{\mu \nu}^a F^{a \mu \nu}
              + (D_{\mu} \Phi)^{\dag} (D^{\mu} \Phi)
%              - \lambda (\Phi^{\dag} \Phi - {v^2 \over 2})^2   \eqno(3)$$
               - V(\phi) \ ,  \eqno (4a)$$
 with
$$  V(\phi) = {\lambda \over 4} ( \phi^2 - {v^2 })^2
       \ , \ \ \ \ \      \phi \equiv \sqrt 2
       (\Phi^{\dag}\Phi)^{1/2}  \ .   \eqno(4b)      $$

Sphalerons are saddle points of the
classical energy functional [8].
In order to construct sphalerons
we choose the gauge $A_0=0$,
and we employ the static spherically symmetric ansatz for the fields
$$  \phi(\vec r\,)  = {v} L(r)\ , \ \ \ \ \ \ \
   A_i^a(\vec r\,) = {G(r) \over g r^2} \epsilon_{iba}r_b  \ . \eqno(5)$$
The classical energy functional then reads
$$ \eqalign {  E = {2 \pi M_W \over g^2}
      \int dx \Bigl[ & {G^2(G-2)^2 \over x^2}
   + 2 ({dG \over dx})^2 + 2 G^2 L^2\cr
                & + 4 x^2 ({dL \over dx})^2  + {8 \lambda \over  g^2} x^2
  (L^2 - 1^2)^2\Bigr] \ , \cr}  \eqno(6)$$
%- {\gamma \over \lambda}\hat T L^3 - V_{\infty}]]
where $x$ is the dimensionless coordinate
$$  x = M_W r  $$
and $M_W$ and $M_H$ are the masses of the gauge and Higgs bosons
$$M_W = {g v \over 2} \ , \ \ \ M_H = v \sqrt{2\lambda} \ . $$
In the calculations we use the values
$v \approx 246$ GeV and $g \approx 0.65$, corresponding
to $M_W=80$ GeV.

In order to have non-trivial regular, finite energy
solutions, the radial functions $G(x)$ and $L(x)$
must obey the boundary conditions
$$  \eqalign{ &G(0) = 2 \ , \ \ \ \ \ \ \ G(\infty) = 0 \ , \cr
              &L(0) = 0 \ , \ \ \ \ \ \ \ L(\infty) = 1 \ . \cr}
    \eqno(7)$$
For the energy functional (6)
with boundary conditions (7)
one saddle point solution is known,
the sphaleron [8], whose energy increases from
$7$ TeV for $M_H = 0$ to $13$ TeV for $M_H =\infty$.
For $M_H < 12 M_W$ the sphaleron has
precisely one direction of instability.
For $M_W > 12 M_W$ new directions of instablilty
of the sphaleron appear, which are associated with
new solutions of the electroweak model.
(These are the bisphalerons [18,19], which are based on
the general spherically symmetric ansatz for the fields
involving three functions for the gauge fields and two
functions for the Higgs field.)

{\bf Sphalerons at finite temperature}

In order to introduce finite temperature
effects into the electroweak model
and to describe the electroweak phase transition,
one has to replace the Mexican hat potential (4b)
by a temperature dependent effective potential [6].
The expression for the energy functional
and the equations of motion are then
modified accordingly.
We now discuss the effects of
three temperature dependent effective potentials
on the sphaleron solution.

{\sl Case I.}

The simplest approximation for the temperature dependent
effective potential consists of supplementing the tree level
potential by the leading term
of the high-temperature expansion [6].
Neglecting logarithmic terms, this effective potential reads [6]
$$ V(\phi , T) =  { {\lambda}\over 4} {\phi}^4
                 - {\lambda \over 2} v^2 {\phi}^2
                 + {{\gamma T^2} \over 2} {\phi}^2 \ , \ \ \ \ \
       \gamma =  {{2 M_W^2 + M_Z^2 + 2 M_t^2}\over 4 v^2}
\ . \eqno(8)   $$
This potential has a transition
between the broken phase for $T < T_c$
and the unbroken phase for $T > T_c$ at
$$ T_c^2  = { \lambda v^2 \over \gamma} \ . \eqno(9)$$
The vacuum expectation value of the Higgs field
$\langle \phi (T)\rangle$
is a continuous function of temperature
$$ \eqalign {
  &\langle \phi (T) \rangle  =
    0 \ \ \ \ \ \ \ \ \ \ \ \ \ \ \ \ \ \ \
    \ \ {\rm for} \ \ T > T_c \ , \cr
  &\langle \phi (T) \rangle  =
    v (1 - {\gamma T^2 \over { \lambda v^2}})^{1/2}
                             \ \ \ \ {\rm for} \ \ T < T_c \ . \cr}
\eqno(10)$$

The corresponding sphaleron solutions possess
a nice property. They can be constructed
from their zero temperature counterpart by a suitable scaling
of $x$ and of $L(x)$
$$ \tilde x = x/a \ ,\ \ \ \ \ \tilde L(x) =  a L(x)
\ , \ \ \ \ \  a^2 = (1 - {\gamma T^2 \over { \lambda v^2}})
\ . \eqno(11)$$
In this case the sphaleron energy $E (\lambda,T)$
is exactly given by formula (3)
$$E(\lambda, T) = E(\lambda, T=0)
{\langle \phi (T) \rangle \over {\langle \phi (0) \rangle }}
\ , $$
where $\langle \phi (0) \rangle \equiv v = 246$ GeV.

Since the phase transition is of second order,
this effective potential leads to a
restauration of the baryon-antibaryon symmetry
in the broken phase shortly after the phase transition.
It can hardly be reconciled with the observed BAU.

{\sl Case II. $\theta_w=0$}

Considering higher orders in the perturbative calculation
one obtains corrections to the
temperature dependent effective potential.
In the next order one finds the so called ``one loop
improved'' potential [7]
$$ V(\phi , T) =
  {{\lambda}\over 4} {\phi}^4 - {\lambda\over 2} v^2 {\phi}^2
    + {\gamma T^2 \over 2} {\phi}^2 - \delta T \phi^3 \ ,
    \ \ \ \ \
    \delta =  {{2 M_W^3 + M_Z^3} \over {4 \pi v^3}}
\ . \eqno(12)$$
The new term, cubic in $\phi$, now renders the phase transition
first order.

For this effective potential there are three relevant critical
temperatures. For low temperatures, the minimum of the
potential is attained at some $\langle \phi (T)\rangle \neq 0$
and $\phi = 0$ corresponds to a local maximum of $V$.
At $T=T_c$ (defined in eq.~(9)) the nature of the extremum
at the origin changes.
$\phi = 0$ turns into a local minimum,
which is separated from
the absolute minimum by a small potential barrier.
The two minima become degenerate at a temperature $T=T_b$,
and $\phi = 0$ becomes the absolute minimum for $T>T_b$.
Thus at high temperatures the symmetry is restored.
The local minimum then disappears at $T= T_a$.
To summarize, the expectation value of the Higgs field
is given by the absolute minimum of the potential
$$\eqalign{& \langle \phi (T)\rangle =
   0 \ \ \ {\rm for} \ \ \ T > T_b \ , \cr
           & \langle \phi (T)\rangle >
   0 \ \ \ {\rm for} \ \ \ T < T_b \ , \cr}
\eqno(10') $$
and one observes a first order phase transition.
Choosing the masses $M_W=M_Z=80$ GeV,
$M_H = 45$ GeV and $M_t = 120$ GeV
the three critical temperatures correspond to
$$          T_c = 0.2902v    \ , \ \ \ \ \
            T_b = 0.2962v    \ , \ \ \ \ \  T_a = 0.2965v  $$
in units of $v = 246$ GeV.

For the ``one loop improved'' potential (12)
the last term in the energy functional (6)
must be replaced by
$$ {8\lambda \over g^2} x^2
  [ L^4 -2 L^2 + {{\gamma T^2} \over {2 \lambda v^2}} L^2
               - {4 \delta T \over {\lambda v}} L^3 + C] \ ,  \eqno(13)$$
where the constant $C$ must be adjusted such
that the potential approaches zero asymptotically.
The boundary conditions for the function $L(x)$ become
$$  L(0) = 0 \ , \ \ \ \ \ \ \
    L(\infty) = {{ \langle \phi (T)\rangle}\over v} \ , \eqno(7')$$
where $\langle \phi (T)\rangle$ denotes the non-trivial minimum
of the potential $V(\phi,T)$.

We have analysed the sphaleron for this effective potential
numerically.
The sphaleron is physically meaningful only for $T<T_b$,
but the solutions can be constructed up to
$T<T_a$.
The profiles of $G(x)$ and $L(x)$
are illustrated in Figs.1 and 2
for the temperatures $T=0$ and $T=T_b$,
using the parameters $M_W=M_Z=80$ GeV,
$M_H = 45$ GeV and $M_t = 120$ GeV.
The energy of the sphaleron is shown in Fig.3 (solid line)
as a function of temperature for the same parameters.
The sphaleron energy obtained for this potential
with the approximation formula (3) is also shown in Fig.3 (dashed line).
Obviously, eq.~(3) constitutes a rather good approximation,
since its values for $E_{\rm sp}(T)$ exceed the exact values
for the potential (12) only slightly,
typically by 4\% at $T=T_c$ up to 8\% at $T=T_b$.

In Fig.4 we present the ratio $E_{\rm sp}(T)/T$ for
the parameters $M_W=M_Z=80$ GeV, $M_H = 45$ GeV and $M_t = 120$ GeV.
For convenience, we have represented
the temperature in Fig.4 via the variable  $\xi$
$$   \xi =   {\lambda \gamma \over {\delta ^2}}(1 - ({T_c \over T})^2)
    \eqno(14)   $$
defined [20] such that the critical temperatures $T_c,T_b,T_a$
correspond to $\xi = 0 , \ \xi = 2 , \ \xi = 9/4$, respectively.
Variation of the top quark mass $M_t$ within the experimental bounds
hardly effects the ratio $E_{\rm sp}(T)/T$.
This observation is easily understood at the critical
temperature $T_c$ by applying the approximation formula (3)
$$ {E_{\rm sp}(T_c)\over T_c} = {E_{\rm sp}(0) \over v}
  {{3 \delta}\over \lambda} \ , $$
which is independent of the top quark mass $M_t$.

The model-independent relation (2) provides
a cosmological upper bound on the Higgs mass,
depending on the effective temperature dependent potential.
In Fig.5 we present the ratio $E_{\rm sp}(T)/T$
at the critical temperature $T_c$
as a function of the Higgs mass for the
``one loop improved'' potential for the parameters
$M_W=M_Z=80$ GeV and $M_t = 120$ GeV.
To satisfy relation (2) the Higgs mass must be smaller
than 46 GeV (practically independent of $M_t$),
which is considerably below
the experimental lower bound on the Higgs mass, $M_H > 60$ GeV.

{\sl Case II. $\theta_w \ne 0$}

So far we have considered the sphaleron in the limit
of vanishing mixing angle (i.~e.~$\theta_w =0$). At zero temperature
this constitutes an excellent approximation to the physical
case (i.~e.~$\theta_w \approx 30^0$). The energy
of the sphaleron at the physical mixing angle is only lower by 1\%.
Computing the first order correction in $\theta_w$
to the sphaleron energy at finite temperature, we observe
that it is also 1\%, if $M_W=M_Z$ is chosen in the effective potential.
Including the mixing angle dependence of the parameters
$\gamma$ and $\delta$
in the effective potential, however, leads to a considerable effect
for the ratio $E_{\rm sp}(T)/T$.
For comparison with the $\theta_w=0$ case we show
in Fig.4 the ratio $E_{\rm sp}(T)/T$
for the masses $M_W=80$ GeV, $M_Z=92$ GeV,
$M_H=45$ GeV and $M_t=120$ GeV.
By including the mixing angle dependence
of the gauge boson masses explicitly,
the cosmological bound on the Higgs mass is slightly improved.
Here it increases to $M_H < 50$ GeV.

In Ref.~[8] it was observed that
the sphaleron carries a large magnetic dipole moment.
In the leading order approximation in $\theta_w$
the magnetic dipole moment is given by
$$    \mu = \lim_{x\rightarrow \infty} 2 x^3 p(x) {e\over \alpha_w M_W}
\ , \eqno(15) $$
where $p(x)$ is detemined by the equation
$$  x^2 p'' + 4 x p' = {{L^2 G \over 2}}   \eqno(16a)$$
and satisfies the boundary conditions
$$   p'(0) = 0 \ , \ \ \ \ \ \ \ p(x \gg 1) \approx {\mu \over {x^3}}
\ . \eqno(16b)$$
Our numerical analysis indicates that the magnetic dipole moment
depends slightly on temperature,
when expressed in units of $e/\alpha_w M_W(T)$.
Employing in the effective potential the parameter set
$M_W=80$ GeV, $M_H=45$ GeV, $M_t=120$ GeV and $M_Z=92$ GeV,
we find for the magnetic moment in these units
%$$   \mu(T=0) = 1.9\ , \ \ \ \ \ \mu(T=T_c) = 4.6
%\ , \ \ \ \ \   \mu(T=T_b) = 6.8\ , \ \  \eqno(17a)   $$
$$   \mu(T=0) = 1.90\ , \ \ \ \ \ \mu(T=T_c) = 1.98 (4.0)
\ , \ \ \ \ \   \mu(T=T_b) = 2.02 (6.1)\ ,   \eqno(17)   $$
(where the values in brackets are in units of $e/\alpha_w M_W(0)$).

{\sl Case III.}

Finally let us briefly discuss the sphaleron
when Debye screening effects
are taken into account in the effective potential.
Assuming $M_W = M_Z$,
the effective potential with Debye screening effects reads [9,21]
$$  \eqalign  {   V_{sc} (\phi,T) =
    {\lambda \over 4} {\phi}^4  - {\lambda \over 2} v^2 {\phi}^2
    +&{T^2 \over 8} {\phi}^2 ({{3 M_W^2 + 2 M_t^2}\over v^2}) \cr
    -&{T \over 4 \pi} \Bigl(
      {\phi}^3  {{2 M_W^3} \over v^3}
                 + ({11 \over 6} g^2 T^2
                 +  {{M_W^2}\over v^2} {\phi}^2  )  ^{3 \over 2}
                          \Bigr) \ . \cr
               } \     \eqno(18)   $$
The effective potentials (12) and (18) differ in their cubic pieces.
Potentials like (18) have also been considered by Khoze [15]
in an attempt to find a potential suitable to describe the
barrier between the vacua and not only the vacua themselves.

The sphaleron energies obtained
for the potential (18) are shown in Fig.3 (solid curve)
along with the energies obtained for the potential (12).
The same parameters have been used for both potentials.
Also for this potential formula (3)
for the sphaleron energy (dashed curve)
represents a good approximation.

At a fixed temperature
the sphaleron obtained for the potential with Debye screening effects
is heavier than the one corresponding to the
``one loop improved'' potential.
But on the other hand the phase transition
for the potential with Debye screening effects occurs
at a higher temperature.
Considering the model-independent relation (2) we find
for the effective potential with Debye sceening
$$ \eqalign{
      &E_{\rm sp}(T)/T \approx 35  \ \ {\rm for}
        \ \  T = T_c \approx 0.3075 \ , \cr
      &E_{\rm sp}(T)/T \approx 25  \ \ {\rm for}
        \ \  T = T_b \approx 0.311 \ . \cr}
\eqno(19)  $$
The ratio $E_{\rm sp}(T)/T$ is shown if Fig.5
for the critical temperature $T_c$
as a function of the Higgs mass for the parameters
$M_W=M_Z=80$ GeV and $M_t = 120$ GeV.
The curve obtained with Debye screening effects
included is distinctly below the curve
for the ``one loop inproved'' potential.
While initially there was hope that potentials
incorporating the Debye screening effects
would allow for a higher cosmological bound
on the Higgs mass [9], these results indicate,
that taking Debye screening effects into account
is not favourable for the bound on the Higgs mass.
Indeed, we find from the model-independent relation (2)
for the potential (18)
a cosmological upper bound on the Higgs mass
$M_H < 40$ GeV as compared to $M_H < 46$ GeV
for the ``one loop improved'' potential (12).

{\bf Conclusions}

In this paper we have presented the sphaleron energy
as a function of temperature for
three effective temperature dependent potentials.
We have demonstrated that the simple scaling formula (3)
represents a good approximation for the
sphaleron energy for the ``one loop improved'' potential
and for the effective potential with Debye screening effects included.
We conjecture, that the formula will also be good
for other effective potentials,
i.~e.~that it is sufficient to know the minimum of the respective
effective potential to obtain a good estimate of its sphaleron
energy.

The model-independent relation (2) provides
a cosmological upper bound on the Higgs mass,
depending on the respective effective potential considered.
For $\theta_w=0$ we have obtained the bounds $M_H < 46$ GeV
for the ``one loop improved'' potential and
the even lower value $M_H < 40$ GeV
for the effective potential with Debye screening
effects included.
We therefore conclude that Debye screening effects
are not favourable for the cosmological upper bound
on the Higgs mass.
In contrast, considering the mixing angle dependence
of the effective potential, when calculating the sphaleron energy,
does shift the upper bound on the Higgs mass to a higher value,
$M_H < 50$ for the ``one loop improved'' potential.

A bound of $M_H < 50$ GeV (resp. $M_H < 46$ GeV) is inconsistent
with the present limit from the LEP experiments,
$M_H > 60$ GeV.
However, the bound on the Higgs mass is sensitive
to the effective potential.
Thus employment of a more sophisticated temperature dependent potential
(than the ones considered here) might reconciliate
the bound (2) with the experimental limit.

\vskip 2 cm
{\bf Acknowledgements}\hfill\break
 Conversations with J. Orloff and M.E.~Shaposhnikov are gratefully
   acknowledged.

\vfill \eject

{\bf References}

\item [1] V.~A. Kuzmin, V.~A. Rubakov and M.~E. Shaposhnikov,
          Phys. Lett. B155 (1985) 36.
\item [2] M.~E. Shaposhnikov, Nucl. Phys. B287 (1987) 757,
 B299 (1988) 797.
\item [3] P. Arnold and L. McLerran, Phys. Rev. D36 (1987) 581;
 D37 (1988) 1020.
\item [4] A. Ringwald, Phys. Lett. B201 (1988) 510.
\item [5] L. Carson, X. Li, L. McLerran and R.-T. Wang,
 Phys. Rev. D42 (1990) 2127.
\item [6] E.~W. Kolb and M.~S. Turner, ``The Early Universe'',
 Addison-Wesley Publishing Company, Redwood City, 1990.
\item [7] M.~E. Shaposhnikov, "Anomalous Fermion Number Non-Conservation"
           \break \noindent     CERN-TH.6304/91 (1991).
\item [8] F.~R. Klinkhamer and N.~S. Manton, Phys. Rev. D30 (1984) 2212.
\item [9] M.~E. Shaposhnikov, Phys. Lett. B 277 (1992) 324.
\item [10] L. Dolan and R. Jackiw, Phys. Rev. D9 (1974) 3320.
\item [11] M.~E. Carrington, Phys. Rev. D45 (1992) 2933.
\item [12] M. Dine, R.~G. Leigh, P. Huet, A. Linde and D. Linde,
          Phys. Rev. D46 (1992) 550.
\item [13] J.~R. Espinosa, M. Quiros and F. Zwirner, Phys. Lett. B291
           (1992) 115.
\item [14] D. Brahm and S. Hsu, CALT-68-1705, CALT-68-1762.
\item [15] V.~A. Khoze, "Comment on Thermal Sphaleron Transitions"
          MIT-CTP 92-2103 (1992).
\item [16] M. Dine, P. Huet and R. Singleton Nucl. Phys. B375 (1992) 625.
\item [17] J. Kunz, B. Kleihaus and Y. Brihaye, Phys. Rev. D46 (1992) 3587.
\item [18] J. Kunz and Y. Brihaye, Phys. Lett. B216 (1989)353.
\item [19] L.~G. Yaffe, Phys. Rev. D40 (1989) 3463.
\item [20] N.~Turok, Phys. Rev. Lett. 68 (1992) 1803.
\item [21] M.~E. Shaposhnikov, private communication.

\vfill\eject
{\bf Figure captions}

\noindent {\bf Figure 1} \hfill \break
\noindent The gauge field function $G(x)$ is shown
for the sphaleron obtained with the potential (12)
with parameters $M_H=45$ GeV, $M_W=M_Z=80$ GeV, $M_t=120$ GeV.
The solid (dashed) line
represents the solution for $T=0$ ($T=T_b$).
\vskip 0.5 cm

\noindent {\bf Figure 2} \hfill \break
\noindent The Higgs field function $L(x)$ is shown
for the sphaleron obtained with the potential (12)
with parameters $M_H=45$ GeV, $M_W=M_Z=80$ GeV, $M_t=120$ GeV.
The solid (dashed) line
represents the solution for $T=0$ ($T=T_b$).
\vskip 0.5 cm

\noindent {\bf Figure 3} \hfill \break
The sphaleron energy $E_{\rm sp}$ (in TeV)
is plotted as a function of temperature $T$ (in units of $v$),
for the potentials (12) and (18).
The solid lines represent the values obtained
by numerically solving the equations of motion,
the dashed lines represent the values obtained
from the simple approximation (3).
\vskip 0.5 cm

\noindent {\bf Figure 4}  \hfill \break
The ratio $E_{\rm sp}(T)/T$ is plotted as a function of the variable $\xi$
(see eq.~(14)) for the potential (12) for
the parameters $M_H = 45$ GeV, $M_W=80$ GeV, $M_Z=80$ GeV (resp.
$M_Z=92$ GeV) and $M_t = 120$ GeV.
\vskip 0.5 cm

\noindent {\bf Figure 5}  \hfill \break
The ratio $E_{\rm sp}(T_c)/T_c$
is plotted as a function of the Higgs mass $M_H$ (in units of GeV)
for the potentials (12) and (18) (solid lines)
for the parameters $M_W=M_Z=80$ GeV and $M_t = 120$ GeV.
The dashed line represents the model-independent relation (2).

\vfill \end